\documentclass[twocolumn,showpacs,preprintnumbers,amsmath,amssymb,superscriptaddress]{revtex4}

\usepackage{graphicx}
\usepackage{epstopdf}
\usepackage{dcolumn}
\usepackage{bm}

\def\be{\begin{equation}}
\def\ee{\end{equation}}
\def\bea{\begin{eqnarray}}
\def\eea{\end{eqnarray}}
\newcommand{\ket}[1]{|#1\rangle}
\newcommand{\bra}[1]{\langle#1|}

\begin{document}

\preprint{APS/123-QED}

\title{A non-degenerate optical parametric oscillator as a high-flux
source for quantum lithography}

\author{Hugo Cable}
\email{cqthvc@nus.edu.sg}
\affiliation{Horace C. Hearne Jr. Institute for Theoretical Physics,
Department of Physics and Astronomy, Louisiana State University, Baton
Rouge, Louisiana 70803, USA}
\affiliation{Centre for Quantum Technologies, National University of
Singapore, 3 Science Drive 2, Singapore 117543}

\author{Reeta Vyas}
\email{rvyas@uark.edu}
\affiliation{Department of Physics, University of Arkansas,
Fayetteville, Arkansas 72701}

\author{Surendra Singh}
\affiliation{Department of Physics, University of Arkansas,
Fayetteville, Arkansas 72701}

\author{Jonathan P. Dowling}
\affiliation{Horace C. Hearne Jr. Institute for Theoretical Physics,
Department of Physics and Astronomy, Louisiana State University, Baton
Rouge, Louisiana 70803, USA}

\date{\today}

\begin{abstract}
We investigate the use of a non-degenerate parametric oscillator (NDPO)
as a source for quantum lithography, for which the light can have
high-flux and strong non-classical features.  This builds on the
proposal of Boto, et al. [A. N. Boto, et al., PRL {\bf 85}, 2733 (2000)],
for etching simple patterns on multi-photon absorbing materials with
sub-Rayleigh resolution, using special two-mode entangled states of
light.  An NDPO has two outgoing modes differentiated by polarization or
direction of propagation, but sharing the same optical frequency. We
derive analytical expressions for the multi-photon absorption rates when
the NDPO is operated below, near, and above its threshold. The resulting
interference patterns are characterized by an effective wavelength half
that for the illuminating modes.  We compare our results with those for the 
case of a high-gain optical amplifier source, 
and discuss the relative merit of the NDPO.
\end{abstract}

\pacs{42.50.-p, 42.65.Yj, 42.50.St}

\maketitle

\section{\label{sec:Intro} Introduction}

The Rayleigh criterion states that diffraction limits the
resolution of a traditional-optical lithographic system, and that the
minimum feature size is determined by half the wavelength $\lambda$ of
the illuminating beam.  This limit arises in part from the fundamental
photon statistics of laser light, according to which the constituent
photons behave as if they are uncorrelated.  In the year 2000,
 Boto, et al., set out a concrete proposal for
exploiting non-classical states of light to achieve resolution
beyond the classical limit (termed super-resolution) in a lithography experiment \cite{Boto00}.
Their proposal has attracted considerable research interest
\cite{Agarwal01,DAngelo01,Boyd05,Fukutake05,Agarwal07,Tsang07,Sciarrino08}.
A key idea is to use a source of path-entangled states of light of the form
$\vert N0\rangle+\vert0N\rangle$, in the photon-number basis,
popularly termed ``$N00N$'' states.  In the scheme,
 the $N00N$ states propagate through a simple
interferometer, and then interfere at a $N$-photon--absorbing 
recording material.  In principle, the procedure can etch a
series of straight lines corresponding to an effective wavelength
$\lambda/\left(2N\right)$.  Variations of the method have been proposed
for creating more general one- and two-dimensional interference
patterns, by employing a family of entangled states \cite{Kok01}.
However, considerable work remains to be done to demonstrate the feasibility  of
the method.  A general program aims to do so, and requires investigations in turn of the
source, imaging system, and multi-photon absorption process.

For the case of two-photon quantum lithography, one could use parametric
down-conversion in a  medium exhibiting a $\chi^{(2)}$
optical non-linearity as a source of photon pairs.  Due to the
Hong-Ou-Mandel effect, a pair of indistinguishable photons
simultaneously incident on a symmetric beam splitter will yield a 2-photon
$N00N$-state at the output ports \cite{Hong87}.  Subsequent quantum
interference of the two modes on a 2-photon absorbing material will then
create a  fringe pattern with half the fringe spacing achieved using a classical light source.
However, the output of common parametric down-conversion experiments is
not expected to be sufficiently bright to be useful for typical 2-photon
recording materials.  In Ref.~\cite{Agarwal01}, Agarwal, et al.,
considered a  strongly-pumped
high-gain optical parametric amplifier.  By assuming a single-mode operation,
it can be shown that the
state generated by an unseeded optical parametric amplifier (OPA) is the
two-mode squeezed vacuum state, which has the form $\sum_{n=0}^\infty
\tanh^n(G)\vert n\rangle\vert n\rangle$ in the basis of photon-number
states.  Here, $G$ is a gain parameter, which depends on the interaction
volume in the crystal, the amplitude of the electric field of the pump
beam, and the strength of the second-order susceptibility $\chi^{(2)}$.
To quantify the contrast of the interference pattern at the output, the
visibility is defined as the difference of the maximum and minimum
absorption rates, divided by sum of these rates.  The visibility varies
 between 0 and 1.  Explicit calculation reveals that the
visibility falls from 1 to an asymptotic value of 0.2, as the gain parameter $G$ of
the OPA is increased from $0 \to \infty$.   This result was
generalized in Ref.~\cite{Agarwal07} to the case of higher-order multi-photon
absorbing materials.  Essentially the same behavior is seen in these
cases, with a halving of the fringe spacing, and the visibility falling
from 1 to an asymptotic value, which is greater the higher the order of
the absorption process.  These predicted features have been
demonstrated experimentally, using coincidence measurements at
photodetectors to simulate the recording medium \cite{Sciarrino08}.

In this paper, we focus on using a NDPO (non-degenerate optical
parametric oscillator) as a practical  high-flux source of
non-classical light for quantum lithography.  In this case, the process
of parametric amplification occurs in an optical cavity, in resonant
signal and idler modes.  The signal and idler modes share the same
optical frequency, but have orthogonal polarization.  The cavity modes
are coupled to external propagating modes through a transmissive
end-mirror.  Photons are created in pairs in the intracavity modes, but
evolve independently out of the cavity on a timescale characterized by
the cavity lifetime.  The twin beams at the output are highly
correlated.   Unlike  the OPA, the NDPO has a well-defined
threshold for oscillation, and the below-, near- and above-threshold
regimes require different mathematical treatments.  The theory for the
NDPO is complicated by the need to account for mode losses, and the
coupling of cavity and external modes.  To analyze the quantum
lithography procedure, we will apply a methodology developed and applied
in a series of papers which investigate the properties of the output
fields of the optical parametric oscillators \cite{Vyas89,Vyas92,Vyas95,Vyas06}.  By using the
positive-$P$ representation
\cite{Drummond80}, the Markovian master equation describing the
dynamics of the standard model of the NDPO can be translated into a
multi-variate Fokker-Planck equation.  The Fokker-Planck equation can be
treated using techniques from nonequilibrium classical statistical
mechanics \cite{Carmichael,GardinerZoller}, allowing steady-state
expectation values of various observables to be evaluated.

In Sec.~\ref{sec:Method}, we describe the setup for a quantum
lithography experiment, and present a quantum-classical
correspondence based on the positive-$P$ representation for  the
 master equation for the density matrix.  A solution of the resulting Fokker-Planck
and Langevin equations is presented following Refs.~\cite{Vyas95,Vyas06}.  In
Sec.~\ref{sec:Results}, we propagate the $c$-number variables for
the NDPO's output modes through a simple interferometer, and determine
the multi-photon absorption of arbitrary order.  We
then look in detail at the features of the interference patterns,
focusing on the below-threshold case in Sec.~\ref{sec:Below},
and the near- and above-threshold case in Sec.~\ref{sec:NearAbove}.
 In Sec.~\ref{sec:Discussion} we compare our results with those for the OPA,
 and comment on their similarities and differences.

\section{\label{sec:Method} Method}

\subsection{\label{sec:Method-Setup} The setup}

\begin{figure}[h]
\includegraphics[width=8cm]{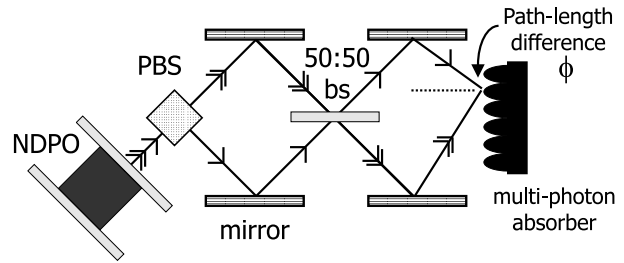}
\caption{\label{fig:setup}
A strongly-pumped NDPO serves as the source, with signal, idler and pump
modes interacting via a $\chi^{(2)}$--- nonlinearity inside an optical
cavity.  Signal and idler modes are frequency degenerate and have
different polarizations.  External signal and idler modes are separated
into different spatial paths by a polarizing beam splitter.  The two
modes are then combined at a 50:50 beam splitter, and are recombined at a
 multi-photon--absorbing recording material.}
\end{figure}

The setup for a quantum lithography experiment is illustrated in
Fig.~\ref{fig:setup}.  We start at the NDPO source.  The modes
corresponding to the signal and idler have frequency $\omega$, and we
denote the corresponding creation operators as $\hat{a}^{\dag}_1$ and
$\hat{a}^{\dag}_2$.  The signal and idler modes are assumed to experience
linear losses characterized by the  decay rate $\gamma$, which arises  from
transmission, absorption, and
scattering losses.  We assume the pump field to be a classical field of
frequency $2\omega$ and of normalized amplitude denoted by $\epsilon$ (assumed to be
given by a positive value).  The cavity pump mode is subject to linear
loss with decay rate $\gamma_3$, and we assume that
$\gamma_3\gg\gamma$.  The creation operator corresponding to the cavity
pump mode is $\hat{a}^{\dag}_3$.  The Hamiltonian describing this system in the
interaction picture can be written as,
\be
\hat{H}_I\!=\!i\hbar \kappa
\left( \hat{a}_{1}^{\dagger }\hat{a}_{2}^{\dagger
}\hat{a}_{3}\!-\!\hat{a}_{1}\hat{a}_{2}\hat{a}_{3}^{\dagger }\right)
+i\hbar \gamma _{3}
\varepsilon \left( \hat{a}_{3}^{\dag }\!-\!\hat{a}_{3}\right)
+\hat{H}_{\text{loss}},\label{hamiltonian}
\ee
where $\kappa$ is the mode-coupling constant determined by the
strength of the optical nonlinearity, and $\hat{H}_{\text{loss}}$ describes the interaction of 
the principal modes with all the
reservoir modes.  Light emitted by the cavity into
the external signal and idler modes, assumed to have orthogonal
polarizations, is separated into two spatial paths by a polarizing beam
splitter.  These modes are then combined at a symmetric beam splitter, and
finally recombined at the multi-photon absorbing material using
mirrors, such that the beams are counterpropagating  at grazing
incidence over the area of interest.  Two classical beams incident on
single-photon absorbing material would generate a fringe pattern of the
form $\propto \left[1\!+\!\cos(2kx)\right]$, where $k$ denotes the optical wave number
for the signal and idler modes, and $x$ denotes translation across the
substrate.  We shall denote the optical phase difference $2kx$ by
$\phi$.  When the recording medium is a $p$-photon absorber, and the
illuminating beams have statistics which need not be classical, the
absorption rate is given by $\sigma^{(p)} {\rm
tr}\left(\hat{\rho}{\hat{E}^{+}(x)}^p{\hat{E}^{-}(x)}^p\right)$, where
$\hat{\rho}$ is  the density matrix for the state of the illuminating field, and $\sigma^{(p)}$
is a generalized cross-section for the process.  This result holds
whenever the optical field may be considered stationary,
quasi-monochromatic and resonant \cite{Mollow68,Agarwal70}. It can be seen that
 the absorption process will  be strongly influenced by
the statistical properties of the light \cite{Qu92}.

\subsection{\label{sec:Method-MasterEq} Master equation and solution
using the positive-$P$ distribution}

Following a standard model for the NDPO, the Markovian
master equation for the reduced density operator for the pump, signal
and idler fields is given by \cite{Carmichael1,GardinerZoller} ,
\bea
\label{eqn:mastereqn}
\frac{\partial \hat{\rho}}{\partial t}&=&\frac{1}{i\hbar }\left[
\hat{H}_{I},\hat{\rho}\right]
+\gamma \sum _{i=1}^{2}\left( 2\hat{a}_{i}\hat{\rho} \hat{a}_{i}^{\dag
}-\hat{a}_{i}^{\dag }\hat{a}_{i}\hat{\rho} -\hat{\rho} \hat{a}_{i}^{\dag
}\hat{a}_{i}\right) \nonumber \\
&& +\gamma _{3}\left( 2\hat{a}_{3}\hat{\rho} \hat{a}_{3}^{\dag
}-\hat{a}_{3}^{\dag }\hat{a}_{3}\hat{\rho} -\hat{\rho} \hat{a}_{3}^{\dag
}\hat{a}_{3}\right),
\eea
where $\hat{H}_I$ here excludes the $\hat H_{loss}$ term of Eq. (\ref{hamiltonian}), which couples the
cavity modes to reservoir modes.
We now map this master equation into a classical
Fokker-Planck equation using the positive-$P$ representation introduced
by Gardiner and Drummond \cite{Drummond80}.  For the current problem this representation
is defined as follows,
\begin{equation}
\label{eqn:posP}
\hat{\rho}\!=\!
\int_{D}\prod_{i=1}^3
d^{2}\alpha _{i}d^{2}\alpha _{i\ast }
\frac{\left\vert \alpha _{i}\right\rangle  \left\langle \alpha _{i\ast
}\right\vert }
{\left\langle \alpha _{i\ast }|\alpha _{i}\right\rangle }
P\left(\vec{\alpha} \right).
\end{equation}
The six variables
$\alpha_1$,$\alpha_{1*},\alpha_2$,$\alpha_{2*},\alpha_3$ and
$\alpha_{3*}$ are independent  complex variables, and we have written
$\vec{\alpha}\!\equiv\!\left(\alpha_1,\alpha_{1*},\alpha_2,\alpha_{2*},\alpha_3,\alpha_{3*}\right)$.
It should be emphasized that the asterisks in the variable indices do
not correspond to complex conjugation.  The  complex variables
$\alpha_i$ and $\alpha_{i*}$ correspond to the mode operators via the relations:
 $\hat{a}_i\ket{\alpha_i}\!=\!\alpha_i\ket{\alpha_i}$ and
 $\bra{\alpha_{i*}}\hat{a}^\dagger_i\!=\!\alpha_{i*}\bra{\alpha_{i*}}$.
The distribution function $P(\vec{\alpha})$ may be assumed to have the
mathematical properties of a probability density function: it is real
valued, positive, and normalized to one when integrated over the full domain
$D$ of $\vec{\alpha}$.  For the master equation,
Eq.~(\ref{eqn:mastereqn}), the corresponding positive-$P$ function
satisfies the following,
\bea
\label{eqn:fullFP}
\frac{\partial P}{\partial t}&\!\!=\!\!&\Big[
\frac{\partial }{\partial \alpha _{1}}\!\left( \gamma \alpha
_{1}\!-\!\kappa \alpha _{2\ast }\alpha _{3}\right) \!+\!
\frac{\partial }{\partial \alpha _{2}}\!\left( \gamma \alpha
_{2}\!-\!\kappa \alpha _{1\ast }\alpha _{3}\right) \!+\! \nonumber \\
&&
\frac{\partial }{\partial \alpha _{1\ast }}\!\left( \gamma \alpha
_{1\ast }\!-\!\kappa \alpha _{2}\alpha _{3\ast }\right) \!+\!
\frac{\partial }{\partial \alpha _{2\ast }}\!\left( \gamma \alpha
_{2\ast }\!-\!\kappa \alpha _{1}\alpha _{3\ast }\right) \!+\! \nonumber \\
&&
\frac{\partial }{\partial \alpha _{3}}\left( \gamma _{3}\alpha
_{3}+\kappa \alpha _{1}\alpha _{2}-\gamma _{3}\varepsilon \right) +
\nonumber \\
&&
\frac{\partial }{\partial \alpha _{3\ast }}\left( \gamma _{3}\alpha
_{3\ast }+\kappa \alpha _{1\ast }\alpha _{2\ast }-\gamma _{3}\varepsilon
\right) + \nonumber \\
&&
\kappa \frac{\partial ^{2}\alpha _{3}}{\partial \alpha _{1}\partial
\alpha _{2}}
+\kappa \frac{\partial ^{2}\alpha _{3\ast}}{\partial \alpha _{1\ast
}\partial \alpha _{2\ast }}
\Big]P.
\eea
This has of the form of a multivariate Fokker-Planck equation \cite{Risken}, and the
use of the positive-$P$ distribution ensures that the diffusion matrix
is positive.  The  Langevin equations corresponding to Eq. (\ref{eqn:fullFP}) can be written as,
\bea
\label{eqn:fullLangevin}
\frac{d\alpha_{1}}{dt}&=&-\gamma \alpha _{1}+\kappa \alpha _{2\ast
}\alpha _{3}+\sqrt{\kappa \alpha _{3}}\frac{\left[ \xi _{1}\left(
t\right)\!+\!i\xi _{2}\left( t\right) \right]}{\sqrt{2}}, \nonumber \\
\frac{d\alpha_{2}}{dt}&=&-\gamma \alpha _{2}+\kappa \alpha _{1\ast
}\alpha _{3}+\sqrt{\kappa \alpha _{3}}\frac{\left[ \xi _{1}\left(
t\right)\!-\!i\xi _{2}\left( t\right) \right]}{\sqrt{2}}, \nonumber \\
\frac{d\alpha_{3}}{dt}&=&-\gamma _{3}\alpha _{3}-\kappa \alpha
_{1}\alpha _{2}+\gamma _{3}\varepsilon, \nonumber \\
\frac{d\alpha_{1\ast }}{dt}&=&-\gamma \alpha _{1\ast }+\kappa \alpha
_{2}\alpha _{3\ast }+\sqrt{\kappa \alpha _{3\ast }}\frac{\left[ \xi
_{3}\left( t\right) \!-\!i\xi _{4}\left( t\right) \right]}{\sqrt{2}},
\nonumber \\
\frac{d\alpha_{2\ast }}{dt}&=&-\gamma \alpha _{2\ast }+\kappa \alpha
_{1}\alpha _{3\ast }+\sqrt{\kappa \alpha _{3\ast }}\frac{\left[ \xi
_{3}\left( t\right) \!+\!i\xi _{4}\left( t\right) \right]}{\sqrt{2}},
\nonumber \\
\frac{d\alpha_{3\ast }}{dt}&=&-\gamma _{3}\alpha _{3\ast }-\kappa \alpha
_{1\ast }\alpha _{2\ast }+\gamma _{3}\varepsilon,
\eea
where the $\xi_i$ are real-valued, white-noise, stochastic variables with
$\langle\xi_j(t)\rangle=0$ and
$\langle\xi_i(t)\xi_j(t')\rangle=\delta_{ij}\delta(t-t')$ for all values
of the indices.

Since it is assumed for the mode loss parameters that
$\gamma_3\gg\gamma$, the pump field can be adiabatically eliminated.
Setting $\dot{\alpha}_3=\dot{\alpha}_{3*}=0$ then leads to,
\bea
\label{eqn:a3a3star}
\alpha _{3}&=&\varepsilon -\frac{\kappa }{\gamma _{3}}\alpha _{1}\alpha
_{2}, \nonumber \\
\alpha _{3\ast }&=&\varepsilon -\frac{\kappa }{\gamma _{3}}\alpha
_{1\ast }\alpha _{2\ast }.
\eea
Substituting Eq.~(\ref{eqn:a3a3star}) into Eq.~(\ref{eqn:fullLangevin}),
we obtain,
\bea
\label{eqn:reducedLangevin}
\frac{d\alpha _{1}}{d\tau }\!\!&=&\!-\alpha _{1}\!+\!\frac{\left(
\sigma\!-\!2\alpha _{1}\alpha _{2}\right)\alpha _{2\ast }}{n_{0}}
\!+\!\sqrt{\frac{\sigma\!-\!2\alpha _{1}\alpha _{2}}{n_{0}} }\left(
\frac{\xi _{1}\!+\!i\xi _{2}}{\sqrt{2}}\right)\!\!,
\nonumber \\
\frac{d\alpha _{2}}{d\tau }\!\!&=&\!-\alpha _{2}\!+\!\frac{\left(
\sigma\!-\!2\alpha _{1}\alpha _{2}\right)\alpha _{1\ast }}{n_{0}}
\!+\!\sqrt{\frac{\sigma\!-\!2\alpha _{1}\alpha _{2}}{n_{0}} }\left(
\frac{\xi _{1}\!-\!i\xi _{2}}{\sqrt{2}}\right)\!\!,
\nonumber \\
\frac{d\alpha _{1\ast }}{d\tau }\!\!\!&=&\!\!\!\!-\alpha _{1\ast
}\!\!+\!\!\frac{\left( \sigma\!\!-\!\!2\alpha _{1\ast }\alpha _{2\ast
}\right) \alpha _{2}}{n_{0}}\!+\!\!\sqrt{\frac{\sigma\!\!-\!\!2\alpha
_{1\ast }\alpha _{2\ast }}{n_{0}} }\left( \frac{\xi _{3}\!-\!i\xi
_{4}}{\sqrt{2}}\right)\!\!, \nonumber \\
\frac{d\alpha _{2\ast }}{d\tau }\!\!\!&=&\!\!\!\!-\alpha _{2\ast
}\!\!+\!\!\frac{\left( \sigma\!\!-\!\!2\alpha _{1\ast }\alpha _{2\ast
}\right)\alpha _{1}}{n_{0}} \!+\!\!\sqrt{\frac{\sigma\!\!-\!\!2\alpha
_{1\ast }\alpha _{2\ast }}{n_{0}} }\left( \frac{\xi _{3}\!+\!i\xi
_{4}}{\sqrt{2}}\right)\!\!, \nonumber \\
\eea
where time  has been scaled in terms of the cavity lifetime $\tau=\gamma t$.
Parameter $n_0=2\gamma\gamma_3/\kappa^2$ is proportional to the square of the number of photons
in the cavity at threshold, and sets the scale for the number of photons necessary to explore the nonlinearity 
of interaction.  $\sigma=2\gamma_3\epsilon/\kappa$ is a
scaled measure of the pump field amplitude relative to its value at threshold.  At threshold the rates
$\gamma$ and $\epsilon\kappa$ are equal, and therefore $\sigma=n_0$.

Using the reparametrization set out in Refs.~\cite{Vyas92,Vyas95}, this
set of Langevin equations can be decoupled, and the dimensionality of
the problem can be reduced from eight to four.  We define four
real-valued parameters $u_1$,$u_2$,$u_3$, and $u_4$ according to,
\be
\label{eqn:changeofvar}
\left(
\begin{array}{c}
  \alpha_1 \\
  \alpha_2 \\
  \alpha_{1*} \\
  \alpha_{2*} \\
 \end{array}
\right)
=
\left(\frac{\sigma^2}{8n_0}\right)^{1/4}
\left(
\begin{array}{cccc}
  1 & 1 & i & i \\
  1 & 1 & -i & -i \\
  1 & -1 & -i & i \\
  1 & -1 & i & -i \\
 \end{array}
\right)
\left(
\begin{array}{c}
  u_1 \\
  u_2 \\
  u_3 \\
  u_4 \\
 \end{array}
\right)\,.\ee
The $u_i$ may be interpreted as scaled pseudo-quadrature variables.
An exact distribution for the positive-$P$ function for the NDPO may now
be written down, as presented first in \cite{Vyas95}, valid for below-,
near-, and above-threshold regimes.  For
$n_0\sim10^6-10^8$, a typical value for  laboratory
systems \cite{Holliday86}, the positive-$P$ function is given to a very good
approximation by
\bea
\label{eqn:fullposP}
P\left(\vec{u}\right)
&\!\!\propto\!\!&
\exp
\Big\{
a_{1}\left( u_{1}^{2}\!+\!u_{3}^{2}\right)\!+\!a_{2}\left(
u_{2}^{2}\!+\!u_{4}^{2}\right)
\nonumber \\
&&
-\frac{1}{2}
\left[
\left( u_{1}^{2}\!+\!u_{2}^{2}\!+\!u_{3}^{2}\!+\!u_{4}^{2}\right) ^{2}
\!+\!4\left( u_{1}^{2}\!+\!u_{3}^{2}\right) \left(
u_{2}^{2}\!+\!u_{4}^{2}\right)
\right]
\Big\}, \nonumber \\
\eea
where $\vec{u}=\left(u_1,u_2,u_3,u_4\right)$.  Parameters $a_1$, $a_2$
correspond to the pump strength and are given by,
\bea
\label{eqn:a1a2def}
a_1&=&\sqrt{2n_0}\left(r-1\right), \nonumber \\
a_2&=&-\sqrt{2n_0}\left(r+1\right),
\eea
where $r=\sigma/n_0=\kappa\epsilon/\gamma$.  It can be seen from Eq. (\ref{eqn:fullposP}) that any
moment of the form,
$$\left\langle
u_{1}^{n_{1}}u_{2}^{n_{2}}u_{3}^{n_{3}}u_{4}^{n_{4}}\right\rangle
\!=\!
\int_{-\infty }^{\infty }\!\int_{-\infty }^{\infty }\!\int_{-\infty
}^{\infty }\!\int_{-\infty }^{\infty }
u_{1}^{n_{1}}u_{2}^{n_{2}}u_{3}^{n_{3}}u_{4}^{n_{4}}P\left( \vec{u}\right),$$
has the value zero if any of the $n_i$ is odd.  The distribution of
Eq.~(\ref{eqn:fullposP}) can be used to evaluate the steady-state
expectation values for any normally-ordered product of the field operators.  In the next
section, we will compute the absorption rates for multi-photon recording
media, exploiting the symmetries of $P\left(\vec{u}\right)$ and making
approximations valid for the different
regimes of operation of the NDPO.

\section{\label{sec:Results} Results}

In this section, we compute the multi-photon absorption rates at the
recording medium for a quantum-lithographic process, by computing
expectation values using the positive-$P$ distribution presented in
Eq.~(\ref{eqn:fullposP}).  We first propagate the signal and idler
variables $\alpha_1,\alpha_{1*}$ and $\alpha_2,\alpha_{2*}$ through the
imaging apparatus, as described in Sec.~\ref{sec:Method}.  By
interfering the signal and idler fields at a symmetric 50:50 beam
splitter, the fields at the output are given by,
\bea
\label{eqn:fieldsthroughbs}
\beta_1&=&\frac{(-\alpha_1+i\alpha_2)}{\sqrt{2}} \nonumber \\
\beta_2&=&\frac{(-\alpha_2+i\alpha_1)}{\sqrt{2}} \nonumber \\
\beta_{1*}&=&\frac{(-\alpha_{1*}-i\alpha_{2*})}{\sqrt{2}} \nonumber \\
\beta_{2*}&=&\frac{(-\alpha_{2*}-i\alpha_{1*})}{\sqrt{2}}.
\eea
Propagating the fields to the multi-photon recording material, the
combined field, at a location corresponding to an
optical phase difference of $\phi$, is given by $\alpha_3=(\beta_1
e^{i\phi}+\beta_2)$ and $\alpha_{3*}=(\beta_{1*}
e^{-i\phi}+\beta_{2*})$.  Substituting Eq.~(\ref{eqn:fieldsthroughbs})
in these expressions, we obtain,
\bea
\label{eqn:fieldatabsorber}
\alpha_3&=&\frac{1}{\sqrt{2}}
\left[
\alpha_1(-e^{i\phi}+i)+\alpha_2(ie^{i\phi}-1)
\right]
\nonumber \\
\alpha_{3*}&=&\frac{1}{\sqrt{2}}
\left[
\alpha_{1*}(-e^{-i\phi}-i)+\alpha_{2*}(-ie^{-i\phi}-1)
\right].
\eea
The rate of $p$-photon absorption then is given by the average quantity,
\be
\label{eqn:absorptionratedef}
I^p_3(\phi)\!=\!\langle\alpha_{3}(\vec{u})^p\alpha_{3*}(\vec{u})^p\rangle.
\ee
Substituting Eq.~(\ref{eqn:changeofvar}), we also have,
\bea
\label{eqn:a3multa3star}
\alpha _{3\ast }\alpha _{3}\!\!\!&=&\!\!\!
\frac{1}{2c^{2}}\left[ \left( u_{1}\!-\!u_{2}\right) \cos \left(
\frac{\phi }{2}\right) \!+\!i\left( u_{3}\!-\!u_{4}\right) \sin \left(
\frac{\phi }{2}\right) \right] \nonumber \\
&& \left[ \left( u_{1}\!+\!u_{2}\right) \cos \left( \frac{\phi
}{2}\right) \!-\!i\left( u_{3}\!+\!u_{4}\right) \sin \left( \frac{\phi
}{2}\right) \right]\!\!.
\eea
We first evaluate the fringe pattern for a one-photon absorber, given by
$I^1_3\left(\phi\right)$.
Inspecting the symmetries of the positive-$P$ function,
Eq.~(\ref{eqn:fullposP}), we see that
symmetries exist between the variables $u_1$ and $u_3$, as well as $u_2$
and $u_4$, so that $P(u_1,u_2,u_3,u_4)\!=\!P(u_3,u_2,u_1,u_4)$ and
$P(u_1,u_2,u_3,u_4)\!=\!P(u_1,u_4,u_3,u_2)$.
It follows immediately that
\be \label{eqn:I3}
I^1_3(\phi)\!=\! r \sqrt{2n_0} \left(\left\langle
u_{1}^{2}\rangle\!-\!\langle u_{2}^{2}\right\rangle\right),
\ee
  and there
is no dependence on $\phi$ either below or above
threshold, and the illumination of the substrate is constant across the
surface in the ensemble-averaged sense.
Next we consider the case of multi-photon absorbing materials.

\subsection{\label{sec:Below} Below threshold regime}

Inspecting the positive-P distribution, Eq.~(\ref{eqn:fullposP}), for
the below threshold case, $\kappa\epsilon\ll \gamma$, we find that
 $a_1$ and $a_2$ are large
negative quantities, and the distribution can be approximated by,
\bea
P\left( \vec{u}\right) \!\!&\approx&\!\!
\left( \sqrt{\frac{|a_{1}|}{\pi }}e^{-\left\vert a_{1}\right\vert
u_{1}^{2}}\right)
\left( \sqrt{\frac{|a_{1}|}{\pi }}e^{-\left\vert a_{1}\right\vert
u_{3}^{2}}\right) \nonumber \\
\!\!&&\!\!
\times
\left( \sqrt{\frac{|a_{2}|}{\pi }}e^{-\left\vert a_{2}\right\vert
u_{2}^{2}}\right)
\left( \sqrt{\frac{|a_{2}|}{\pi }}e^{-\left\vert a_{2}\right\vert
u_{4}^{2}}\right).\label{Pbelow}
\eea
For a typical NDPO with $n_0\sim10^6$ this approximation is very good
for the parameter $r$ ranging from $0$ to $0.99$.
The four variables $u_i$ are Gaussian and independent. Their  even order moments
are given by
\bea
\left\langle u_{1}^{2k}\right\rangle &=&\left\langle u_{3}^{2k}\right\rangle= \frac{\left( 2k\right)
!}{k!\left(4\left\vert a_{1}\right\vert \right)^{k} }\;,
\nonumber \\
\left\langle u_{2}^{2k}\right\rangle &=& \left\langle u_{4}^{2k}\right\rangle=\frac{\left( 2k\right)
!}{k!\left(4\left\vert a_{2}\right\vert \right)^{k} }\;,\label{uaverage}\eea
and their odd order moments vanish.

In order to compute the multi-photon absorption rates
$I_3^p\left(\phi\right)$, defined by Eq.~(\ref{eqn:absorptionratedef})
and Eq.~(\ref{eqn:a3multa3star}), a suitable
grouping of terms must be found.  To this end we define the function,
\be
\label{eqn:Fdef}
F\left( u_{i},u_{j},\phi \right) =\left[ u_{i}\cos \left( \phi /2\right)
-iu_{j}\sin \left( \phi /2\right) \right] ^{2},
\ee
so that $I_3^p$ can be expressed as
\bea
\label{eqn:belowthresholddecomp}
 I_{3}^{p}\left( \phi \right)
\!\!\!\!&=&\!\!\!\!
\left( r \sqrt{2n_0}\right) ^{p}\sum_{k=0}^{p}\binom{p}{k}\left(
-1\right) ^{p-k} \nonumber \\
&&
\times\left\langle F\left( u_{1},u_{4},\phi \right) ^{k}\right\rangle
\left\langle F\left( u_{2},u_{3},\phi \right) ^{p-k}\right\rangle\!.
\eea
We exploit the fact that the $u_i$ are not coupled, to evaluate the
moments of separable sums independently.  Since the pairs of variables
$u_1$ and $u_3$, and  $u_2$ and $u_4$,
share the same distribution, we also conclude
\bea
\left( -1\right) ^{p-k}\left\langle F\left( u_{2},u_{3},\phi \right)
^{p-k}\right\rangle& =&\left\langle F\left( u_{3},u_{2},\phi +\pi \right)
^{p-k}\right\rangle \nonumber\\  &=&\left\langle F\left( u_{1},u_{4},\phi +\pi \right)
^{p-k}\right\rangle.\nonumber\eea
Expanding an arbitrary moment $k$ of $F\left( u_{1},u_{4},\phi \right)$
we find
\bea
\left\langle F\left( u_{1},u_{4},\phi \right) ^{k}\right\rangle
\!&=&\!\sum_{l=0}^{k}\binom{2k}{2l}\left\langle u_{1}^{2l}\right\rangle
\left\langle u_{4}^{2k-2l}\right\rangle \left( -1\right) ^{\left(
k-l\right) } \nonumber \\
&& \times \cos ^{2l}\left( \phi /2\right) \sin ^{2k-2l}\left( \phi
/2\right).
\eea
By substituting the expressions for the even moments $u_1$ and $u_4$
given above, and the form of parameters $a_1$ and $a_2$ defined by
Eq.~(\ref{eqn:a1a2def}), this sum can be
evaluated as,
\be
\left\langle F\left( u_{1},u_{4},\phi \right)
^{k}\right\rangle\!=\!\frac{\left( 2k\right) !}{k!4^{k}\left(
\sqrt{2n_{0}}\right) ^{k}\left( 1-r^{2}\right) ^{k}}\left[ r+\cos \left(
\phi \right) \right] ^{k}.
\ee
Using these  relations, we arrive at a
compact expression for the $p$-photon absorption rate,
\bea
\label{eqn:ratebelowthreshold}
I_3^p(\phi)
&=&
\left[ \frac{r}{4\left( 1-r^{2}\right) }\right]
^{p}p!\sum_{k=0}^{p}\frac{\left( 2k\right) !\left( 2p-2k\right)
!}{k!^{2}\left( p-k\right) !^{2}} \nonumber \\
&& \times \left[ r+\cos \left( \phi \right) \right] ^{k}\left[ r-\cos
\left( \phi \right) \right] ^{\left( p-k\right) }.
\eea

\begin{figure}[t]
\fbox{
\includegraphics[width=5cm]{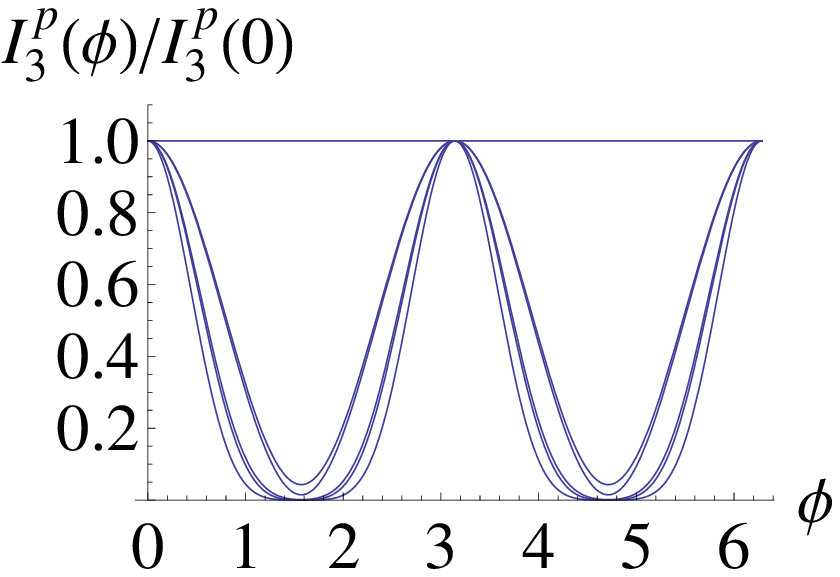}
(a)
}
\fbox
{
\includegraphics[width=5cm]{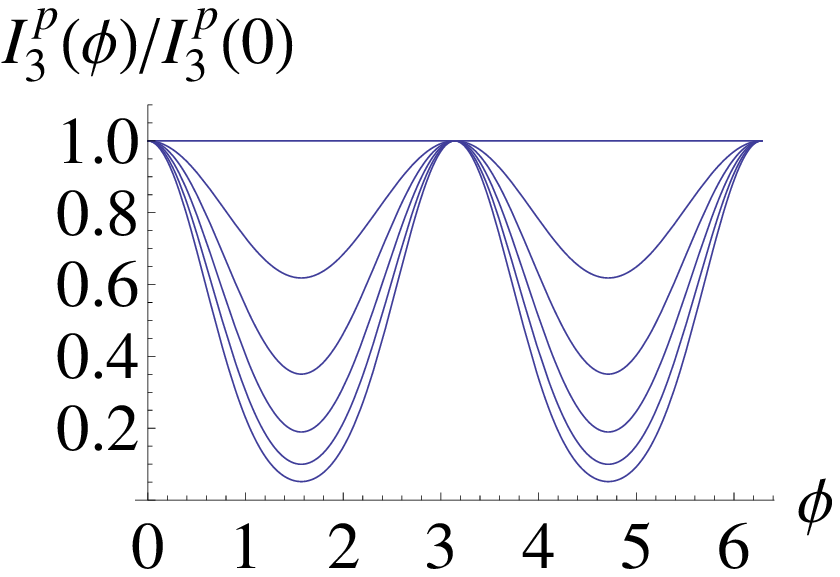}
(b)
}
\caption{\label{fig:belowfringepatterns}
Plots (a) and (b) show the absorption rates $I_3^p(\phi)$, scaled
between 0 and 1,  for $p=1,\cdots,6$.  It is assumed that the NDPO is
operated below threshold, with pump parameter $r=0.15$ in (a) and $r=0.9$ in
(b).  As $p$ increases the corresponding fringe patterns become better
defined.}
\end{figure}
Looking in detail at the form of the absorption rates $I_3^p(\phi)$ in
Eq.~(\ref{eqn:ratebelowthreshold}) we see for the following.  For the
$p=1$ case, the absorption rate is constant and no fringe pattern is
created, as already seen following Eq.~(\ref{eqn:I3}).  For the case of $p\geq2$, $I_3^p(\phi)$ is
given as a sum of even powers of $\cos(\phi)$.  The corresponding fringe
pattern  therefore has  terms with periods corresponding to $\cos(2n\phi)$ where $2n\le p$.  As $p$ increases, there is a greater contribution
from higher power terms with sharper interference patterns.  The fringe
patterns for the cases of $p=1,\cdots,6$ are plotted in
Fig.~\ref{fig:belowfringepatterns}(a) and (b), for the pump parameter $r= 0.15$ and $0.9$
respectively.

\begin{figure}[t]
\fbox{
\includegraphics[width=5cm]{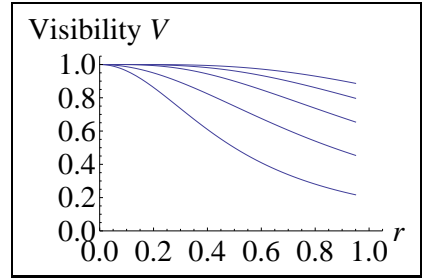}
}
\caption{\label{fig:belowvisibilities}
The visibility, defined by $V=\left[I^p_3(\phi_{\rm
max})\!-\!I^p_3(\phi_{\rm min})\right]/
\left[I^p_3(\phi_{\rm max})\!+\!I^p_3(\phi_{\rm min})\right]$ and
normalized between 0 and 1, is plotted for $p=2,\cdots,6$
as a function of pump parameter $r$ for
 the NDPO operating below threshold.}
\end{figure}
Fig.~\ref{fig:belowvisibilities} shows fringe visibilities  as $r$
ranges from 0 to 0.95.  It is seen that for every case with $p\geq2$,
the visibilities are close to 1 when
the NDPO is operated far below threshold, but as the pump power is
increased the visibilities fall steadily.  In a lithographic process, it
is possible to compensate for unwanted constant exposure by using a
substrate with greater depth, and visibilities greater than 0.2 are
often considered adequate in practice.  This criterion is satisfied in
all cases considered here with $p\geq2$.
Making a comparison with the case of an OPA source reported in the
earlier work of Agarwal, et al., in Ref.~\cite{Agarwal07}, we find the
following.  Eq.~(\ref{eqn:ratebelowthreshold}) is listed
explicitly in Table~\ref{tab:belowabsrates} of
Appendix~\ref{appendix:Below} for the cases of $p=1,\cdots,6$.
Inspecting Eqs.~(19-22) of Ref.~\cite{Agarwal07} we find the formulae
for the multi-photon absorption rates identical on identifying $r$ in
current analysis and $\tanh(G)$ in the analysis of Agarwal, et al., for
which $G$ represents the single-pass gain for the OPA.  The range for
parameter $r$ for the NDPO below threshold  (from $0 \to 1_-$), corresponds to
values of parameter $G$ for the OPA from $0 \to \infty $.
Visibilities for the sub-threshold NDPO are listed in
Table~\ref{tab:belowvisibilities} of Appendix~\ref{appendix:Below} for
$p=1,\cdots,6$, which  agree with the results reported for
the OPA across the corresponding range for the parameter  $G$.

\subsection{\label{sec:NearAbove} Near and above threshold regime}

As the NDPO  is pumped more strongly and passes through threshold,
pump rate $r$ exceeds unity, the nature of the
positive-$P$ distribution changes significantly.  Inspecting the governing parameters
$a_1$ and $a_2$, defined by Eq.~(\ref{eqn:a1a2def}), we see that
while $a_2$  continues to take  large negatives values of increasing size as
$r$ increases, $a_1$ is zero at threshold and takes positive values for
$r>1$.  The positive-$P$ distribution, given by
Eq.~(\ref{eqn:fullposP}), may now be well approximated by the expression below,
which is valid both near and far above threshold for values of $n_0$
greater than $10^4$ (\cite{Vyas06}),
\bea
\label{eqn:posPabovethreshold}
P\left( \vec{u}\right)
\!\!\!&\simeq&\!\!\!
\left( \sqrt{\frac{\left\vert a_{2}\right\vert }{\pi }}e^{-\left\vert
a_{2}\right\vert u_{2}^{2}}\right)
\left( \sqrt{\frac{\left\vert a_{2}\right\vert }{\pi }}e^{-\left\vert
a_{2}\right\vert u_{4}^{2}}\right) \nonumber \\
\!\!\!&\times &\!\!\!
\left( N e^{-\frac{1}{2}\left( u_{1}^{2}+u_{3}^{2}-a_{1}\right)
^{2}}\right),
\eea
where $N$ denotes a normalization factor for the $u_1$,$u_3$ component,
defined by
$$N\!=\!\frac{\sqrt{2}}{\left[\pi ^{3/2}{\rm erfc}\left(
-a_{1}/\sqrt{2}\right)\right]},$$
 and ${\rm
erfc(z)}\equiv( {2}/{\sqrt{\pi }})\int_{z}^{\infty }e^{-s^{2}}ds$ is
the complementary error function.
It can be seen that variables $u_2$ and $u_4$ are independent Gaussian
variables with pump parameter  $a_2$ in the below-threshold regime.
However, variables $u_1$ and $u_3$ with pump parameter $a_1$ are now strongly coupled.  As a
consequence, the decomposition given by
Eq.~(\ref{eqn:belowthresholddecomp}), used to evaluate $I_p(\phi)$ in
the below-threshold case, can no longer be applied.

To proceed,  in this case we look in detail at the moments for the $u_i$ that arise in
computing $I_p(\phi)$ above threshold.
The moments for the Gaussian variables $u_2$ and $u_4$ are as before
given by Eq. (\ref{uaverage}). By reparametrizing $u_1$ and
$u_3$ in polar coordinates,
it follows that,
\bea
\left\langle u_{1}^{2s}u_{3}^{2t}\right\rangle \!
\left\langle u_{2}^{2m}\right\rangle
\!
\left\langle u_{4}^{2n}\right\rangle
\!\!&=&\!\!
R\left( 2 s\!+\!2 t\!+\!1,a_{1}\right)
B\left( s\!+\!\frac{1}{2},t\!+\!\frac{1}{2}\right)
 \nonumber \\
\!\!& \times&\!\!
2N
\frac{\left( 2m\right) !}{m!\left( 4\left\vert a_{2}\right\vert \right)
^{m}}
\frac{\left( 2n\right) !}{n!\left( 4\left\vert a_{2}\right\vert \right)
^{n}}.
\label{uiave}\eea
$B\left(\cdot,\cdot\right)$ here denotes the  Beta function,
defined by $B\left( S,T\right)\equiv\Gamma \left( S\right) \Gamma \left(
T\right) /\Gamma \left( S+T\right)$, and arises from
the integration over the angular component associated with $u_1$ and
$u_3$.  The integration over the remaining radial component leads to the
$R\left(\cdot\right)$ contribution, defined by, $R\left( S,a_{1}\right)
\equiv\int_{0}^{\infty }\rho ^{S}e^{-\frac{1}{2}\left( \rho
^{2}-a_{1}\right) ^{2}}d\rho$. Writing,
\bea
R\left( 2s\!+\!2t\!+\!1,a_{1}\right)
\!\!&=&\!\! \int_{0}^{\infty }\!\!\!\rho ^{2s+2t-1}\left( \rho
^{2}\!-\!a_{1}\right) e^{-\frac{1}{2}\left[ \rho ^{2}-a_{1}\right]
^{2}}d\rho \nonumber \\
\!\!&+& a_{1}R\left( 2s\!+\!2t\!-\!1,a_{1}\right),
\eea
for $s+t\geq1$, and integrating by parts,
\begin{align}
& \int_{0}^{\infty }\rho ^{2s+2t-1}\left( \rho ^{2}\!-\!a_{1}\right)
e^{-\frac{1}{2}\left[ \rho ^{2}-a_{1}\right] ^{2}}d\rho \nonumber\\
&=\begin{cases}\frac{1}{2}e^{-\frac{1}{2}a_{1}^{2}} \,\,\,& (\rm{if}\,\,
s+t=1),\\
\left( s\!+\!t\!-\!1\right) R\left( 2s\!+\!2t\!-\!3,a_{1}\right)
\,\,\, &(\rm{if} \,\, s+t>1),\end{cases}
\end{align}
we get a recursion relation for $R\left(\cdot\right)$.  Here we list first  few radial functions,
\bea
\label{eqn:Rfirstfewcases}
R\left( 1,a_{1}\right)\!\!\!&=&\!\!\!\frac{1}{2\pi N} \nonumber \\
R\left( 3,a_{1}\right)\!\!\!&=&\!\!\!\frac{1}{2}\exp \left(
-\frac{1}{2}a_{1}{}^{2}\right) +a_{1}\frac{1}{2\pi N} \nonumber \\
R\left( 5,a_{1}\right)\!\!\!&=&\!\!\!\frac{a_{1}}{2}\exp \left(
-\frac{1}{2}a_{1}{}^{2}\right) +(1+a_{1}{}^{2})\frac{1}{2\pi N} \nonumber \\
R\left( 7,a_{1}\right)\!\!\!&=&\!\!\!\left(
1+\frac{1}{2}a_{1}^{2}\right) \exp \left( -\frac{1}{2}a_{1}{}^{2}\right)
+\left( 3a_{1}+a_{1}^{3}\right) \frac{1}{2\pi N} \nonumber \\
R\left( 9,a_{1}\right)\!\!\!&=&\!\!\!\left(\!
\frac{5}{2}a_{1}\!+\!\frac{1}{2}a_{1}{}^{3}\!\right) \exp \left(\!
-\frac{1}{2}a_{1}{}^{2}\!\right)\!+\! \frac{\left(
3\!+\!6a_{1}{}^{2}\!+\!a_{1}{}^{4}\right)}{2\pi N} \nonumber \\
\eea
Below threshold ($r=\kappa\epsilon/\gamma < 1$), the parameter $a_1$ is negative.   It is  zero at threshold ($r=\kappa\epsilon/\gamma = 1$), and positive above threshold ($r=\kappa\epsilon/\gamma > 1$).   As $a_1$ increases from negative to positive values the NDPO goes through a phase transition, and the intensities of the signal and idler modes increase very rapidly.  For a typical value of $n_0=10^6-10^8$, the region for phase transition is very narrow and a small change in $r$ from $0.99 \to 1.01$ changes the statistics of the NDPO dramatically \cite{Vyas95,Vyas06}.   In this region, we find that the visibility for $p$-photon absorption changes from the below-threshold to the above-threshold behavior.

\begin{figure}[t]
\fbox{
\includegraphics[width=6.5cm]{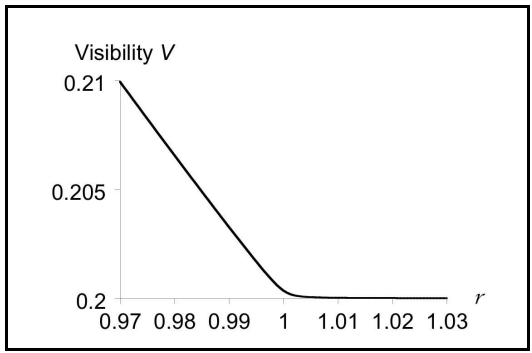}
}
\caption{\label{fig:threshvisibilities}
The visibility for $p=2$   versus  pump parameter $r$ as operation of the NDPO changes from below threshold to the above threshold regime for $n_0=10^6$.}
\end{figure}

 For  the NDPO
operating much  above threshold ($a_1>> 1$),  we can considerably simplify the expression for $I_3^p(\phi)$.  It follows from
Eqs. (\ref{uiave}) and (\ref{eqn:Rfirstfewcases}) that
$\left\langle u_{1}^{2s}u_{3}^{2t}\right\rangle\simeq
B\left(s+1/2,t+1/2\right) a_1^{s+t}/\pi$.  Since powers of $n_0$ appear
only in the denominator for moments of $u_2$ and $u_4$, and in the
expansion of $I_p(\phi)$ the powers of the variables $u_i$ satisfy
$s+t+m+n=p$ , it follows that all contributions from variables $u_2$ and
$u_4$ can be neglected.  Within this approximation,
\be
\alpha _{3}^{p}\alpha _{3\ast }^{p}\simeq\left(r \sqrt{2n_0}\right)
^{p}\left[ u_{1}^{2}\cos ^{2}\left( \phi /2\right) +u_{3}^{2}\sin
^{2}\left( \phi /2\right) \right] ^{p}, \nonumber
\ee
and we find the expectation value,
\bea
\label{eqn:rateabovethreshold}
 I_p\left(\phi\right)\!\!\!\!&=&\!\!\!
 \! \left( r \sqrt{2n_0}\right) ^{p}\!\!\;\;
 \sum_{s=0}^{p}\binom{p}{s}
\left\langle u_{1}^{2s}u_{3}^{2p-2s }\right\rangle
 \nonumber \\  && \times \cos
^{2s}\!\left( \frac{\phi}{2}\right) \sin ^{2p-2s }\!\left(
\frac{\phi}{2}\right)
 \nonumber \\
&=&
 \left( \frac{a_{1}r\sqrt{2n_0}}{8}\right) ^{p}
 \sum_{s=0}^{p} \frac{\left( 2s\right) !\left( 2p-2s\right)
!}{s!^{2}\left( p-s\right) !^{2}}
 \nonumber \\
 && \,\,\, \times \,\,\,
 \left[ 1+\cos \left( \phi \right) \right] ^{s}\!\left[ 1-\cos \left(
\phi \right) \right] ^{p-s}.
\eea
The $p$-photon absorption rates for much above threshold, are
therefore seen to generate fringe patterns
of a form independent of the strength of the pump.  Fig. \ref{fig:threshvisibilities} shows how  the fringe visibility changes near the threshold regime for the 2-photon
case.  Note that as $r$ changes from $0.97 \to 1.03$ the parameter $a_1$ changes  from $-42 \to +42$. As expected, much above threshold the visibility is almost constant.

\section{\label{sec:Discussion} Discussion}

In conclusion, we have found that a NDPO source can be used to generate
fringe patterns with an effective wavelength
half that for the signal and idler modes which interfere at the
recording media, when a $p$-photon absorption
process is available with $p\geq2$.  These fringe patterns have high
visibility, falling from $1$ at low pump power, to asymptotic values of
$0.2$ or greater at high pump power.  Above threshold, the forms of the
fringe patterns are insensitive to the strength of the pump,
whereas below threshold, the forms of the fringe
patterns depend on the pump power, and tend towards the asymptotic case
as the NDPO threshold is approached.  Comparing with the
results reported in Ref.~\cite{Agarwal07}, we find that the fringe
patterns generated using an NDPO operating below threshold  ($r$ from $0\to 1_-$) are similar
to  fringe pattern    generated by an OPA operated with a  corresponding gain ($G$ from $0\to \infty$) .  In the
latter case there is no cavity and no above threshold regime, and the process of optical parametric
amplification occurs in propagating modes.
This result is perhaps
surprising, since the earlier analysis of Ref.~\cite{Agarwal07}
disregards photon losses, and uses a single-mode analysis
by ignoring all but one down converted spatial mode.
The NDPO source has some experimental advantages.  The signal and idler
modes at the outputs of the NDPO are collimated because of the use of a
cavity, and higher-powers can be obtained than is the case for the OPA,
 both of which are important in light of the small
cross-sections for typical multi-photon absorption processes.
Collimated, high-power, outputs also increase the speed at which a
substrate may be imaged --- a critical factor in the mass production of,
say, computer chips.
Finally, the reparameterization of the dynamics of the NDPO in terms of the
four pseudo-quadrature variables $u_i$ above, sheds some light on why
the change in fringe patterns and visibilities are observed.
Far-below threshold, all four of the $u_i$ contribute approximately
equally to the absorption process, and behave as independent
parameters.  As the pump power is increased, and threshold is
approached, parameter $a_1$ tends to $0$.  Since $a_1$ appears in the
denominator of each of the moments of variables $u_1$ and $u_3$, $u_1$
and $u_3$ make a growing contribution compared to $u_2$ and $u_4$.
Above threshold, variables $u_1$ and $u_3$ make the primary contribution
to the absorption process, and the effects of $u_2$ and $u_4$ can be
neglected.  Variables $u_1$ and $u_3$ are also strongly coupled in this
regime.  The analysis of this paper also contributes compact general
formulae for the multi-photon absorption rates,
Eq.~(\ref{eqn:ratebelowthreshold}) and Eq.~(\ref{eqn:rateabovethreshold}).

\begin{acknowledgments}
JPD and HC would like to acknowledge the Army Research Office, the
Defense Advanced Research Projects Agency, and the Intelligence Advanced
Research Projects Activity.  HC further acknowledges support for this
work by the National Research Foundation and Ministry of Education,
Singapore.
\end{acknowledgments}

\newpage
\appendix

\section{\label{appendix:Below} Fringe patterns below threshold}

Based on the general formula for the absorption rates below threshold,
Eq.~(\ref{eqn:ratebelowthreshold}), the following tables list explicitly
the absorption rates, and the visibilities for the corresponding fringe
patterns, for $p$-photon absorbers with $p$ ranging from 1 to 6.

\begin{table}[h]
\caption{\label{tab:belowabsrates} $p$-photon absorption rates as a
function of the optical-path-length difference
$\phi$ and the pump parameter $r$, varying from far below to near
threshold (from 0 to approximately 0.99).}
\begin{ruledtabular}
\begin{tabular}{c|c}
$p$ & $I_3^p(\phi)$ \\
\hline
$1$ & $\frac{r^{2}}{\left( 1-r^{2}\right) }$ \\
$2$ & $\frac{r^{2}}{\left( 1-r^{2}\right) ^{2}}\left[ \cos ^{2}(\phi
)+2r^{2}\right] $ \\
$3$ & $\frac{3r^{4}}{\left( 1-r^{2}\right) ^{3}}\left[ 3\cos ^{2}(\phi
)+2r^{2}\right]$ \\
$4$ & $\frac{3r^{4}}{\left( 1-r^{2}\right) ^{4}}\left[ 3\cos ^{4}(\phi
)+24r^{2}\cos ^{2}(\phi )+8r^{4}\right] $ \\
$5$ & $\frac{15r^{6}}{\left( 1-r^{2}\right) ^{5}}\left[ 15\cos ^{4}(\phi
)+40r^{2}\cos ^{2}(\phi )+8r^{4}\right] $ \\
$6$ & $\frac{45r^{6}}{\left( 1-r^{2}\right) ^{6}}\left[ 5\cos ^{6}(\phi
)+90r^{2}\cos ^{4}(\phi )+120r^{4}\cos ^{2}(\phi )+16r^{6}\right] $ \\
\end{tabular}
\end{ruledtabular}
\end{table}

\begin{table}[h]
\caption{\label{tab:belowvisibilities} The visibilities, defined by
$V=\left[I^p_3(\phi_{\rm max})\!-\!I^p_3(\phi_{\rm min})\right]/
\left[I^p_3(\phi_{\rm max})\!+\!I^p_3(\phi_{\rm min})\right]$ and
normalized between 0 and 1, for the fringe patterns with a $p-$photon
absorbing recording material.
The third column lists the limiting values for the visibilities as the
pump parameter $r$ ranges from far below to near threshold (from 0 to
approximately 0.99).}
\begin{ruledtabular}
\begin{tabular}{c|c|c}
\,\,\,\,\,\,\,\,\,\,\,\,\,\,\,\,\,\,\,\,$p$\,\,\,\,\,\,\,\,\,\,\,\,\,\,\,\,\,\,\,\,
& Visibility $V(r)$ &$ V(0^+)\rightarrow V(1^-)$ \\
\hline
$1$ & $0$ & --- \\
$2$ & $1-\frac{4r^{2}}{\left[ 1+4r^{2}\right] }$ & $1\rightarrow 0.20$ \\
$3$ & $1-\frac{4r^{2}}{\left[ 3+4r^{2}\right] }$ & $1\rightarrow 0.43$ \\
$4$ & $1-\frac{16r^{4}}{\left[ 3+24r^{2}+16r^{4}\right] }$ &
$1\rightarrow 0.63$ \\
$5$ & $1-\frac{16r^{4}}{\left[ 15+40r^{2}+16r^{4}\right] }$ &
$1\rightarrow 0.77$ \\
$6$ & $1-\frac{32r^{6}}{\left[ 5+90r^{2}+120r^{4}+32r^{6}\right] }$ &
$1\rightarrow 0.87$ \\
\end{tabular}
\end{ruledtabular}
\end{table}


\newpage
\bibliography{CABLE_NDPO4QL_sub1}

\end{document}